\RequirePackage{fix-cm}
\documentclass[twocolumn,epjc3]{svjour3}  
\smartqed  
\RequirePackage{graphicx}
\RequirePackage{url}
\RequirePackage{xspace}
\journalname{Eur. Phys. J. C}
\begin{document}

\newcommand{\cov}[2]{\ensuremath{\mathrm{cov}[#1,#2]}}
\newcommand{\covmat}{\ensuremath{\mathcal{M}}}
\newcommand{\dif}[1]{{\mathrm{d}#1\,}}
\newcommand{\pb}{\ensuremath{\mathrm{pb}}\xspace}
\newcommand{\TeV}{\ensuremath{\mathrm{TeV}}\xspace}
\newcommand{\GeV}{\ensuremath{\mathrm{GeV}}\xspace}
\newcommand{\VL}{\ensuremath{V_{L}}\xspace}
\newcommand{\VR}{\ensuremath{V_{R}}\xspace}
\newcommand{\gL}{\ensuremath{g_{L}}\xspace}
\newcommand{\gR}{\ensuremath{g_{R}}\xspace}
\newcommand{\eft}{EFT{\it fitter}\xspace}
\newcommand{\fL}{\ensuremath{f_{L}}\xspace}
\newcommand{\fR}{\ensuremath{f_{R}}\xspace}
\newcommand{\fz}{\ensuremath{f_{0}}\xspace}

\title{\eft---A tool for interpreting measurements in the context of effective field theories
}


\author{
  Nuno Castro\thanksref{addr1}$^{,}$\thanksref{addr2} \and
  Johannes Erdmann\thanksref{addr3} \and
  Cornelius Grunwald\thanksref{addr3} \and
  Kevin Kr{\"o}ninger\thanksref{addr3} \and
  Nils-Arne Rosien\thanksref{addr4}
}



\institute{
Laborat\'orio de Instrumenta\c c\~ao e F\'{\i}sica Experimental de
Part\'{\i}culas, Departamento de F\'{\i}sica, Universidade do Minho,
4710-057 Braga, Portugal \label{addr1} \and
Departamento de F\'{\i}sica e Astronomia, Faculdade de Ci\^encias,
Universidade do Porto, 4169-007 Porto, Portugal\label{addr2} \and
Lehrstuhl f{\"u}r Experimentelle Physik IV, TU Dortmund, Otto-Hahn-Stra{\ss}e~4a, 44227~Dortmund, Germany\label{addr3} \and
II. Physikalisches Institut, Universit{\"a}t G{\"o}ttingen, Friedrich-Hund-Platz~1, 37077~G{\"o}ttingen, Germany\label{addr4}
}

\date{Received: date / Accepted: date}

\maketitle

\begin{abstract}
Over the past years, the interpretation of measurements in the context
of effective field theories has attracted much attention in the field
of particle physics. We present a tool for interpreting sets of
measurements in such models using a Bayesian ansatz by calculating the
posterior probabilities of the corresponding free parameters
numerically. An example is given, in which top-quark measurements are
used to constrain anomalous couplings at the $Wtb$-vertex.  
\end{abstract}


\section{Introduction}
\label{sec:introduction}

With the recent start of the Run-2 of the LHC, searches for physics
beyond the Standard Model (BSM) will reach unprecedented
sensitivity. The LHC's increased centre-of-mass energy opens a new
kinematic regime and enhances the direct production cross section of
heavy, yet unknown, particles---if they exist. It is a good time for
bump hunters.

On the other hand, it is not obvious that the mass scale of such new
particles is anywhere near the energy which can be reached by current,
or even future, accelerators. However, in contrast to the direct
production of heavy particles, their impact on observables accessible
at current collider experiments can be probed indirectly in the
context of effective field theories. Such theories extend the Standard
Model (SM) Lagrangian by terms allowed in quantum field theory and
which share the gauge symmetries of the SM. These terms contain one or
several effective operators and corresponding coefficients, often
referred to as Wilson coefficients, which define the individual
strength of these operators. Depending on the type of operator, the
additional terms in the Lagrangian can have an impact on different
observables, which, in turn, can be compared to a set of corresponding
measurements. Comparisons of SM predictions and observations can be
used to constrain the Wilson coefficients by propagation of
uncertainty. The strategy to indirectly infer on the parameters of a
physics model, may it be an effective model or a full model, has
proven to be successful in a variety of applications, e.g. in the
field of flavor physics~\cite{Beaujean:2013soa},
super-symmetry~\cite{Bechtle:2004pc}, or electroweak precision
measurements~\cite{Flacher:2008zq}.

This paper describes a generic tool, the \eft, for performing such
interpretations in the context of user-defined physics models and
formulating them in terms of Bayesian reasoning. Emphasis is placed on
the statistical treatment of the combination of measurements
correlated by their uncertainties as well as on often overlooked
issues in interpretations, such as the necessity to consider
model-specific efficiency and acceptance corrections for the
measurements, or the presence of physical constraints on observables
and parameters. An example is given for the case of an effective field
theory in the top-quark sector, which is an active field of
research. This example is motivated by the wealth of experimental data
delivered by the Tevatron and LHC experiments and the increasing
precision of measurements involving top quarks. Also from a
theoretical perspective, the interpretation of top-quark measurements
is attractive: recent calculations, e.g. predictions of the cross
section of top-quark pair
production~\cite{Czakon:2012pz,Czakon:2011xx,Czakon:2013goa}, reach
next-to-next-to-leading order (NNLO) precision in perturbative QCD. A
historical example for the interpretation of experimental data in the
context of top quarks is the successful prediction of the top-quark
mass from electroweak precision measurements, see, e.g.,
Ref.~\cite{Langacker:1991an}.

This paper is organised as follows. Section~\ref{sec:measurements}
describes the statistical procedure of combining several measurements,
while their interpretation is discussed in
Section~\ref{sec:interpretation}. The numerical implementation of the
\eft is introduced in Section~\ref{sec:implementation}, and an example
for interpretations in the field of top-quark physics is given in
Section~\ref{sec:example}. The paper is concluded in
Section~\ref{sec:conclusions}.

\section{Combination of measurements}
\label{sec:measurements}

In Bayesian reasoning, inference of the free parameters
$\vec{\lambda}$ of a model $M$ is based on the posterior probability
of those parameters given a data set $\vec{x}$, $p(\vec{\lambda} |
\vec{x})$. It is calculated using the equation of Bayes and
Laplace~\cite{Bayes:1764vd},
\begin{eqnarray}
\label{eqn:posterior}
p(\vec{\lambda} | \vec{x}) = \frac{p(\vec{x} | \vec{\lambda}) \cdot p(\vec{\lambda})}{p(\vec{x})} \, ,
\end{eqnarray}
where $p(\vec{x} | \vec{\lambda})$ is the probability of the data, or
likelihood, and $p(\vec{\lambda})$ is the prior probability of the
parameters $\vec{\lambda}$. In Bayesian literature, the normalisation
constant in the denominator,
\begin{eqnarray}
p(\vec{x}) = \int \dif{\vec{\lambda}} \, p(\vec{x} | \vec{\lambda}) \cdot p(\vec{\lambda}) \, ,
\end{eqnarray}
is often referred to as the evidence.

In the following, we distinguish two types of models: i) those for
which the parameters can be directly measured from the data, and ii)
those for which this is not the case. Models of type ii) typically
predict values of physical quantities, or observables, $\vec{y}$,
which depend on the parameters of the model, $\vec{\lambda}$, so that
$\vec{y} = \vec{y}(\vec{\lambda})$. For example, the predicted cross
sections in scattering processes depend on the couplings and masses of
the particles described by physics models. These couplings and masses
are often not predicted by the models themselves, in which case they
are free parameters. Couplings, e.g., can often only be estimated
indirectly from the measurements of cross sections and other
observables. This case will be discussed further in
Section~\ref{sec:interpretation}. On the other hand, models of type i)
are often used for the plain combination of measurements in which the
physical quantities themselves, e.g. cross sections, angular
distributions or branching ratios, are interpreted as model
parameters, i.e., $\vec{y}=\vec{\lambda}$. We will discuss this case
in the following.

\subsection{Combination of measurements}
\label{sec:combination}

Following the notation of Ref.~\cite{Valassi:2003mu}, we assume to
have $N$ observables, $y_{i}~(i=1,\dots,N)$, which are estimated based
on $n$ measurements, $x_{i}~(i=1,\dots,n)$. Each quantity $y_{i}$ is
measured $n_{i}\ge 1$ times, so that $n=\sum_{i=1}^{N} n_{i} \ge
N$. We adopt the common assumption that the likelihood terms in
Eqn.~(\ref{eqn:posterior}), $p(\vec{x}|\vec{y})$, have a multivariate
Gaussian shape, and that the uncertainties of the measurements of
$x_{i}$ can be correlated. The elements of the symmetric and
positive-semidefinite covariance matrix are
\begin{eqnarray}
\covmat_{ij} = \cov{x_{i}}{x_{j}} \, . 
\end{eqnarray}
Assuming $M$ different sources of uncertainty, the covariance matrix
can be decomposed into contributions from each source,
\begin{eqnarray}
\cov{x_{i}}{x_{j}} = \sum_{k=1}^{M} \mathrm{cov}^{(k)}[x_{i}, x_{j}] \, .
\end{eqnarray}
The likelihood can then be expressed as
\begin{eqnarray}
\label{eqn:likelihood}
-2 \ln p(\vec{x} | \vec{y}) = \sum_{i=1}^{n} \sum_{j=1}^{n}  \left[ \vec{x} - U \vec{y} \right]_{i} \covmat_{ij}^{-1} \left[ \vec{x} - U \vec{y} \right]_{j} \, ,
\end{eqnarray}
where the elements $U_{ij}$ of the $n \times N$-matrix $U$ are unity
if $x_{i}$ is a measurement of the observable $y_{j}$, and zero
otherwise.

The best linear unbiased estimator
(BLUE)~\cite{Valassi:2003mu,Lyons:1988rp} can be found by minimising
the expression in Eqn.~(\ref{eqn:likelihood}), while an estimator in
the Bayesian approach is constructed by inserting
Eqn.~(\ref{eqn:likelihood}) into the RHS of
Eqn.~(\ref{eqn:posterior}), and by specifying prior probabilities for
the parameters. The prior probabilities can include physical
constraints, e.g. the requirements that cross-sections can only take
positive values or that branching ratios lie between zero and
one. Prior knowledge can also come from auxiliary measurements or
theoretical considerations. It is worth noting that combined estimates
of physical quantities based on individual posterior probabilities
need to be cleaned from the corresponding prior probabilities,
i.e. the prior information about a parameter should only be included
once in the overall combination. It should also be noted that prior
probabilities, and in particular physical constraints, can lead to a
strong non-Gaussian shape of the resulting posterior probability
distribution, even if the input measurements are assumed to be
described by Gaussian probability densities.

Typical estimators are the set of parameters which maximise the
posterior probability, the mean values of the posterior probability
distribution, or the set of parameters which maximise the marginal
probabilities,
\begin{eqnarray}
p(y_{i}|\vec{x}) = \int \prod_{j \neq i} \dif{y_{j}} \, p(\vec{y}|\vec{x}) \, .
\end{eqnarray}
For uniform prior probabilities, and in the absence of further
constraints, the global mode of the posterior corresponds to the BLUE
solution~\cite{Lyons:1988rp}. The uncertainty on $y_{i}$ can be
defined as the central interval containing 68\% probability, the set
of smallest intervals containing 68\% probability or simply the
standard deviation of the marginalised posterior. These three measures
are equal for Gaussian distributions. Similarly, simultaneous
estimates of the uncertainties on $y_{i}$ and $y_{j}$ can be obtained
by the two-dimensional contours of the smallest intervals containing,
e.g., 39\% or 68\% probability.~\footnote{The classical one-sigma
  contour contains 39\% probability while the 68\% contour is
  typically shown in the field of particle physics~\cite{brandt}.}
Upper and lower limits on $y_{i}$ are typically set by calculating the
90\% or 95\% quantiles of the corresponding marginal posterior
distribution.

\subsection{Uncertainties of the correlation}

Although it is often straightforward to obtain estimates of the
quantities $\vec{y}$ and of their uncertainties, it is not trivial to
quantify the correlation induced by the different sources of
uncertainties. If, e.g., sources of systematic uncertainty have an
impact on several of those measurements, the correlation is often
assumed to be extreme ($\rho=\pm1$). On the other hand, correlated
statistical uncertainties caused by partially overlapping data sets
are often estimated using pseudo data: two measurements, $x_{i}$ and
$x_{j}$, are obtained from common sets of simulated data and the
linear correlation coefficient
$\rho_{ij}=\cov{x_{i}}{x_{j}}/\sigma_{i}\sigma_{j}$ between the
estimates is calculated. Here, $\sigma_{i}$ and $\sigma_{j}$ are the
standard deviations of $x_{i}$ and $x_{j}$, respectively.

If no reliable estimate of the correlation is possible, one can
associate correlation coefficients with nuisance parameters
$\vec{\nu}$ and choose suitable prior probabilities for these
parameters. Necessary requirements for the priors are that the
correlation coefficients are constrained to be in the interval
$[-1,+1]$, and that the covariance matrix remains
positive-semidefinite for all possible values of $\vec{\nu}$. They
should, however, parameterise the prior knowledge about the
correlation, e.g. by restricting the correlation coefficients to be
positive or by favouring mild correlations. Depending on the problem,
it can be difficult to formulate such priors analytically. In
particular, it is advisable to not allow values resulting in
correlation coefficients of $\rho=\pm 1$ for the entries of the total
covariance matrix as those might lead to numerical
instabilities. Prior probabilities for covariance matrices are
proposed in the literature, see
e.g. Refs.~\cite{Leonard1992,Barnard2000,Huag2013}. The covariance
matrix \covmat, and thus the likelihood, are then functions of
$\vec{\nu}$, so that $p(\vec{x} | \vec{y}, \, \vec{\nu})$. The
prior probability can be factorised, $p(\vec{y}, \, \vec{\nu}) =
p(\vec{y}) \cdot p(\vec{\nu})$, if $\vec{y}$ and $\vec{\nu}$ are
assumed to be independent, which is typically the case. In order to
obtain a function depending only on $\vec{y}$ alone, all nuisance
parameters are integrated out,
\begin{eqnarray}
p(\vec{y}|\vec{x}) = \int \dif{\vec{\nu}} \, p(\vec{y}, \, \vec{\nu}|\vec{x}) \, ,
\end{eqnarray}
and the estimates of $\vec{y}$ are obtained as before from the
resulting marginal posterior probabilities.

\subsection{Propagation of uncertainty}

In cases where the probability density for a quantity $f(\vec{y})$ is
needed, the uncertainty on $\vec{y}$ needs to be propagated to
$f$. For Gaussian posterior probabilities, the propagation of
uncertainty is often done using the well-known rules for uncertainty
propagation. These imply that the function $f$ can be linearised in
$\vec{y}$ and that the posterior probability for the quantity of
interest also has a Gaussian shape. Since this is not always the case,
and since the posterior of the combination does not have to be a
Gaussian due to the additional prior information, we instead propose
to use a numerical evaluation of the uncertainty: if it is possible to
sample from the posterior distribution $p(\vec{y}|\vec{x})$, one can
calculate the target quantity $f(\vec{y})$ for each sampled point. The
obtained frequency distribution for $f$ converges to the posterior
probability distribution $p(f|\vec{x})$ in the large sample limit. As
discussed in Section~\ref{sec:implementation}, such sampling can be
done on-the-fly when using Markov Chain Monte Carlo (MCMC).

\subsection{Ranking the impact of individual measurements}
\label{sec:ranking1}

One is often interested in how much a single measurement contributes
to a combination. In BLUE averages, the contributions are added
linearly using weights: the larger the weight, the more important the
measurement. Although counter-intuitive at first sight, these weights
can also take negative values, induced by strong negative
correlations, see e.g. the discussion in
Refs.~\cite{Valassi:2003mu,Lyons:1988rp}.

There are several ways to estimate the impact of individual
measurements in a combination. For the approach discussed here, we
propose to repeat the combination while removing one measurement from
the combination at a time. The measurements can then be ranked
according to the resulting increase in uncertainty compared to the
uncertainty of the overall combination. For combinations with two
($n$) physical quantities, the area (volume) of the smallest contour
(hyper-sphere) covering, e.g., 68.3\% of the posterior probability,
can be used as a rank indicator. The motivation for this choice is to
answer the question how the combination would change if a particular
measurement had not been considered in the combination.

Regardless of the choice of rank indicator, the estimators themselves,
their uncertainties and the correlations between the physical
quantities should be monitored during this procedure. Note that the
uncertainty can also decrease if particular measurements are removed,
as in the case of outliers. \\

\subsection{Ranking the impact of individual sources of uncertainty}
\label{sec:ranking2}

The situation is similar when one aims to rank the sources of
uncertainties in order of their importance. We propose to repeat the
combination while removing one source of uncertainty from the
combination at a time. The sources of uncertainty can then be ranked
according to the resulting decrease of uncertainty compared to the
uncertainty of the overall combination. The rank indicator can be
extended to $n$-dimensional problems. From a practical point of view,
this approach allows answering the question which source of
uncertainty is most important to improve in future iterations of the
measurements.

\section{Interpretation of measurements in physics models}
\label{sec:interpretation}

Estimating the parameter values $\vec{\lambda}$ of a complex physics
model $M$ based on (the combination of) measurements and the
subsequent propagation of uncertainties is an inverse problem which is
often ill-posed and in most cases difficult to solve.  We propose here
to re-formulate the problem discussed in the previous section in the
following way: instead of directly identifying the observables with
the fit parameters, we instead fit the free parameters of the physics
model under study based on the relation between the observables and
the parameters. If the model predicts observables $y_{i}$ for each set
of parameter values $\vec{\lambda}$, $y_{i}=y_{i}(\vec{\lambda})$,
these are then compared to the measurements $x_{i}$ using a
multivariate Gaussian model. The same formalism as in
Section~\ref{sec:combination} can be used to estimate $\vec{\lambda}$
for a given data set. The likelihood of the model is
\begin{eqnarray*}
p(\vec{x}|\vec{\lambda}) = \int \dif{\vec{y}}  \, p(\vec{x}|\vec{y}) \cdot p(\vec{y}|\vec{\lambda}) \, ,
\end{eqnarray*}
where
$p(\vec{y}|\vec{\lambda})=\delta(\vec{y}-\vec{y}(\vec{\lambda}))$. It
is worth noting that one has to formulate prior probabilities in terms
of the model parameters and not in terms of the measurements
themselves. Physical constraints can be incorporated into the model
predictions, `external knowledge' can be viewed as an additional
measurement.~\footnote{It is a virtue of the Bayesian formalism that
  an update of knowledge is trivially obtained by defining results
  from previous measurements or other considerations as prior
  probabilities of a new measurement.}

Using the same framework also helps to include measurements and
physical quantities in the analysis which would otherwise be combined
separately, e.g. by a working group concerned with cross sections and
another one interested in angular distributions. However, due to
common sources of systematic uncertainties and overlapping data sets,
the posterior probability of the two measurements shows a
correlation---a fact that needs to be considered in the global fit to
a physics model.

\section{Implementation}
\label{sec:implementation}

The \eft~\footnote{The code is available at
  \url{https://github.com/tudo-physik-e4/EFTfitterRelease} and
  includes the code for the example discussed in this paper.} builds
on the Bayesian Analysis Toolkit (BAT)~\cite{Caldwell:2008fw} which is
a software package written in C++ that allows the implementation of
statistical models and the inference on their free
parameters.~\footnote{The BAT code is available at
  \url{http://mppmu.mpg.de/bat/}.}  Several numerical algorithms can
be used to perform the combination and interpretation steps introduced
in this paper: marginal distributions can be calculated using, for
example, Markov Chain Monte Carlo, while global optimisation can be
done using the Minuit implementation of ROOT~\cite{Brun:1997pa}.

\subsection{Definition of a model and observables}

The key component of the \eft is the user's definition of a model. It
is simply characterised by a set of free parameters, e.g. couplings
and masses, and by the predictions of observables as a function of the
model's parameters. Note that the model is not automatically derived
from a user-defined Lagrangian and it is thus not constrained to a
particular class of models. As a consequence, however, the predictions
from the model are not required to be consistent, and it is the user's
responsibility to formulate a meaningful set of predictions. Of
course, interfaces to more complex software tools can be included in
the model.

\subsection{Input}

The input to the EFT\textit{fitter} is a set of measurements including
a break-down of the uncertainties into several categories,
e.g. statistical uncertainties and different sources of systematic
uncertainties.  In addition, the correlation between the measurements
for each category of uncertainty needs to be provided.

It is worth noting that the different measurements need to be unified
in a sense that the sources of uncertainties are treated in the same
categories throughout all measurements, e.g. uncertainties related to
the reconstruction of objects in a collider experiment should have one
or more well-defined categories: uncertainties on the luminosity,
efficiencies or acceptances should each have their own categories,
etc.

The measurements can be any measurable quantity, most commonly a cross
section, a mass or a branching ratio.  It can also be an unfolded
spectrum, in which case each bin is treated as an individual
measurement and the full unfolding matrix is needed as an input. The
unfolding matrix provides the acceptances and efficiencies, which can
be treated consistently in the EFT\textit{fitter} as described in
Sec.~\ref{sec:acceptance}.

All of these inputs are provided in a configuration file in xml
format. The file contains information about the number of observables,
the number of measurements of these observables, the number of
uncertainty categories and the number of nuisance parameters to be
used in the fit.  The allowed range for each observable has to be
provided. For each measurement, the name of the observable, its
measured value as well as its uncertainty in each uncertainty category
have to be provided as well. Furthermore, measurements can be omitted
from the fit by flagging them as inactive. The correlation matrix is
provided in the same configuration file. The correlation coefficients
between measurements can be treated as nuisance parameters in the
fit. For each nuisance parameter, the measurements that the
correlation coefficient refers to as well as its prior probability
have to be specified.

Apart from the measurements, the user needs to specify prior
probability densities for each model parameter.  These can be chosen
freely, e.g. uniform, Gaussian or exponentially decreasing
functions. Physical boundaries for parameters and observables are
addressed by the definition of the prior probabilities and the
predictions for observables, respectively.  A second configuration
file in xml format is used to specify the prior probability densities
for the parameters, their minimum and maximum values, as well as their
SM predictions.

Sometimes it is necessary to include external input in a combination,
e.g. measurements from low-energy observables, $b$-physics or
cosmology. These can either be treated as additional measurements or
as priors on the parameters, depending on the type of information
provided. In both cases, it is usually difficult to estimate the
correlation between the different inputs and the user should carefully
consider whether the choice of correlation made has a strong effect on
the interpretation of the data.

\subsection{Treatment of acceptance and efficiency corrections}
\label{sec:acceptance}

A problem often encountered when measurements are interpreted in terms
of BSM contributions, is that the acceptances and efficiencies may be
different for different BSM processes, while measurements are mostly
performed assuming the acceptances and efficiencies of SM
processes. An example is the measurement of the cross section for the
pair production of a particle, which in BSM scenarios might also be
produced from an additional resonance decay. The acceptance times
efficiency for the BSM process may be different than for the SM
process if the measurement requires a minimum momentum for the
final-state particles.  These requirements may be more frequently
fulfilled in the BSM scenario if the resonance has a very high
mass. It is also clear that the acceptance and efficiency may depend
on the parameters of the BSM theory, such as the mass of the resonance
in this example.

The EFT\textit{fitter} addresses this problem by separating
observables, measurements and parameters, so that the acceptance and
efficiency can be corrected for when comparing the prediction for the
observables with the measurements. For the SM process, this acceptance
and efficiency correction is equal to unity, but for BSM models, the
correction may differ from one and it is, in general, a function of
the parameters of the theory.

\subsection{Output}

A brief summary of the fit results is provided in a text file, while
four figure files are provided for a more detailed analysis of the fit
results.  In one figure file, all one-dimensional and two-dimensional
marginalised distributions of the fitted parameters are saved.  In two
figure files, the estimated correlation matrix of the parameters and a
comparison of the prior and posterior probability density functions
for the different parameters are shown, respectively.  A last figure
file shows the relations between the parameters of the model and the
observables as defined by the underlying model.  An additional text
file provides the post-fit ranking of the measurements determined as
described in Sections~\ref{sec:ranking1} and~\ref{sec:ranking2}.

\subsection{Structure of the implementation}

The code is structured such that a new folder needs to be created for
a specific analysis. Folders are provided for the example discussed in
this paper (Sec.~\ref{sec:example}) as well as a blank example a user
can start from. Each such folder contains a subfolder for the input
configuration files and an empty folder for the result files. It also
contains a rather generic run file which holds the main function for
the run executable. The model specific details, such as the relation
between the parameters and the observables, the acceptance and
efficiency correction etc. are implemented in a class inheriting from
\texttt{BCMVCPhysicsModel}. In this class, it is necessary to provide
in particular a concrete implementation of the virtual method
\texttt{CalculateObservable(\dots)}.  The code is then compiled using
a Makefile, which is provided. A README file contains further details
for users to get started.

\section{An example for effective field theory involving top quarks}
\label{sec:example}

As an example for applications of the \eft, we discuss the constraints
on anomalous top-quark couplings from two sets of observables: the
measurement of the polarisation of $W$-bosons produced in top-quark
decays and the measurements of the $t$-channel top- and antitop-quark
cross sections. Similar cases have been discussed in
Refs.~\cite{AguilarSaavedra:2006fy,AguilarSaavedra:2008gt,AguilarSaavedra:2008zc,Zhang:2010dr,Buckley:2015nca,Buckley:2015lku,Bernardo:2014vha,Fabbrichesi:2014wva,Birman:2016jhg}. The
concrete model we use is that of Ref.~\cite{AguilarSaavedra:2006fy},
where the Lagrangian describing the $Wtb$-vertex does not only include
a purely left-handed coupling with relative strength \VL, but also a
right-handed vector coupling with strength \VR as well as left- and
right-handed tensor couplings \gL and \gR. The generalised Lagrangian
then takes the form
\begin{eqnarray}
\label{eqn:lagrangian}
\mathcal{L} & = & -\frac{g}{\sqrt{2}}\bar{b}\gamma^{\mu} \left( \VL P_{L} + \VR P_{R} \right) t W_{\mu}^{-} - \\ 
 & & \frac{g}{\sqrt{2}} \bar{b} \frac{i \sigma^{\mu\nu}q_{\nu}}{M_{W}} \left( \gL P_{L} + \gR P_{R} \right)  t W_{\mu}^{-} + h.c.\, ,
\end{eqnarray}
where $g$ is the weak coupling constant, $P_{L}$ and $P_{R}$ are left-
and right-handed projection operators and $M_{W}$ is the mass of the
$W$-boson. In the SM, the left-handed coupling strength is given by
$\VL=\left| V_{tb}\right|^{2}\approx 1$, while the three other
coupling strengths are $\VR=\gL=\gR=0$.

The physical model defined by the Lagrangian in
Eqn.~(\ref{eqn:lagrangian}) has four free parameters: \VL, \VR, \gL,
\gR. For simplicity, we assume these parameters to be real.

\subsection{Observables and predictions}

The observables described in the following are calculated based on
Refs.~\cite{AguilarSaavedra:2006fy,AguilarSaavedra:2008gt,Bach:2014zca}. The
masses of the top quark, the $W$-boson and the bottom quark are
assumed to be $172.5\,\GeV$, $80.4\,\GeV$ and $4.8\,\GeV$,
respectively.

\begin{figure}[t]
  \begin{center}
    \includegraphics[width=0.45\textwidth]{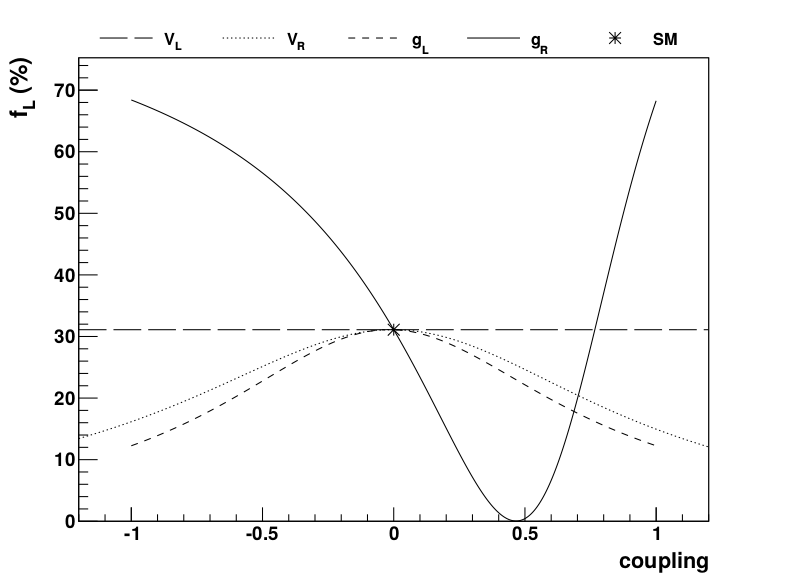} 
    \caption{Fraction of left-handed $W$-bosons from top-quark decays
      as a function of anomalous couplings. The asterisk indicates the
      SM values.
      \label{fig:observables_fl}}
  \end{center}
\end{figure}

The $W$-bosons produced in top-quark decays can be left-handed,
right-handed and longitudinally polarised. The fraction of events with
either of these polarisations are \fL, \fR and \fz, and often referred
to as \emph{helicity fractions}. At NNLO accuracy in the strong
coupling, these fractions are predicted to be $\fL=0.311 \pm 0.005$,
$\fR=0.0017 \pm 0.0001$ and $\fz=0.687 \pm 0.005$ in the
SM~\cite{Czarnecki:2010gb}. Assuming the Lagrangian defined in
Eqn.~(\ref{eqn:lagrangian}), these fractions are functions of the four
coupling strengths. As an example, Fig.~\ref{fig:observables_fl} shows
the predicted fraction of left-handed $W$-bosons as a function of any
of the couplings in the range $[-1,+1]$, while keeping the other three
couplings fixed to their SM values. While the variation of \VR and \gL
results in variations of \fL from about 15\% to a maximum of roughly
30\% at $\VR=0$ and $\gL=0$, \gR has a stronger impact. The value of
\fL ranges from 0\% to about 70\% with a minimum at approximately
$\gR=0.5$. As expected, there is no dependence of the helicity
fractions on \VL. Since the sum of the three fractions is unity, only
two of the three, \fL and \fz, are considered as well as their
correlation.

\begin{figure}[t]
  \begin{center}
    \includegraphics[width=0.45\textwidth]{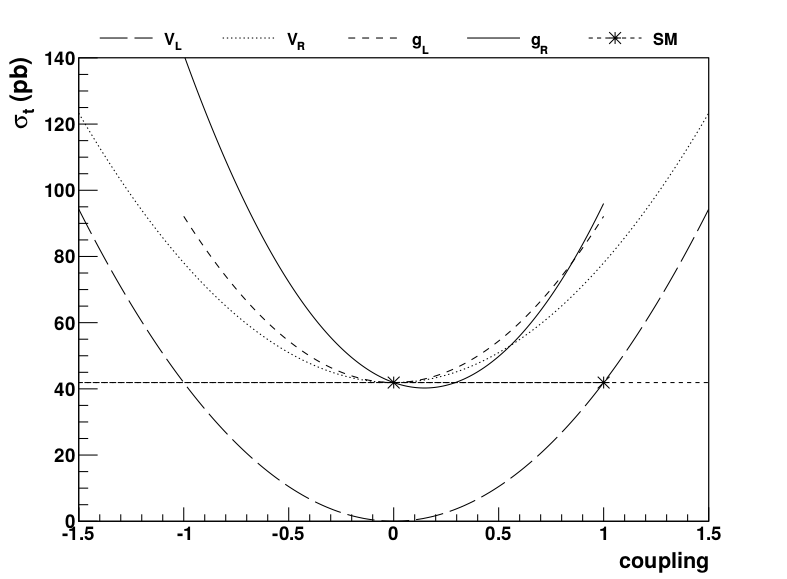} 
    \caption{Cross section for $t$-channel single top-quark production
      as a function of anomalous couplings. The asterisk indicates the
      SM values.
      \label{fig:observables_xs}}
  \end{center}
\end{figure}

Similarly, the cross sections for single top- and antitop-quark
production in the $t$-channel at a centre-of-mass-energy of
$\sqrt{s}=7\,\TeV$ are predicted in the SM to be
$\sigma_{t}=41.9^{+1.8}_{-0.8}\,\pb$ and
$\sigma_{\bar{t}}=22.7^{+0.9}_{-1.0}\,\pb$ at NNLO accuracy in the
strong coupling~\cite{Kidonakis:2011wy}. As an example, the single
top-quark cross section as a function of the four coupling strengths
is illustrated in Fig.~\ref{fig:observables_xs}, again fixing the
other couplings to their SM values for each of the curves. All four
couplings change the predicted cross section resulting in values of
$\sigma_{t}$ between 0~\pb and 140~\pb, where a minimum can be found
at coupling values of about 0.

\subsection{Measurements and assumptions}

The measurements of the $W$-boson polarisation and the $t$-channel
cross sections considered in this example are taken from
Refs.~\cite{Chatrchyan:2013jna} and~\cite{Aad:2014fwa},
respectively. The uncertainties on these measurements are assumed to
be derived from multivariate Gaussian distributions. The measured
values are
\begin{eqnarray*}
\fL & = & 0.310 \pm 0.031 \\
\fz & = & 0.682 \pm 0.045 \\
\sigma_{t} & = & 46 \pm 6 \, \pb \\
\sigma_{\bar{t}} & = & 23 \pm 3 \, \pb \, .
\end{eqnarray*}
The correlation between the measurements of \fL and \fz is quoted in
the reference as $\rho(\fL,\fz)=-0.95$. We assume that there is no
correlation between the helicity and cross-section measurements. This
simplifying assumption is made because the measurements are performed
by two different experiments and because the event selections for the
two measurements are orthogonal. Common sources of systematic
uncertainty, e.g. from LHC machine settings or the modelling of top
quarks in Monte Carlo generators, could lead to a small correlation,
however. Furthermore, we assume that the correlation between the
measurements of top- and antitop-quark cross sections is mild, but not
negligible, and we thus choose
$\rho(\sigma_{t},\sigma_{\bar{t}})=+0.50$. Although the event
selections are again orthogonal---the leptons selected differ by the
sign of their electric charge---common sources of systematic
uncertainty have a similar impact on both measurement,
e.g. uncertainties on the jet-energy and lepton-momentum scales or the
Monte Carlo generator uncertainties. The total uncertainties and the
full correlation matrix used are shown in Tab.~\ref{tab:covariance}.

\begin{table}[ht!]
  \caption{Total uncertainties and the correlation matrix of the
    helicity fraction and $t$-channel cross section measurements.
    \label{tab:covariance}}
  \begin{center}
    \begin{tabular}{llllll}
      \hline
      & Uncertainty & \fL & \fz & $\sigma_{t}$ & $\sigma_{\bar{t}}$ \\
      \hline 
      \fL              & 0.031 & \phantom{-}1.00   & -0.95 & 0.00 & 0.00 \\
      \fz              & 0.045 & -0.95   & \phantom{-}1.00 & 0.00 & 0.00 \\
      $\sigma_{t}$     & 6\,pb    & \phantom{-}0.00 & \phantom{-}0.00 & 1.00 & 0.50  \\
      $\sigma_{\bar{t}}$ & 3\,\pb  & \phantom{-}0.00 & \phantom{-}0.00 & 0.50 & 1.00  \\
      \hline
    \end{tabular}
  \end{center}
\end{table}

The efficiency times acceptance of the measured $t$-channel cross
section depends on the anomalous couplings assumed because they have
an impact on the kinematic distributions of the final-state
particles. Since its evaluation would require further studies
including simulations of the detector setup and a repetition of the
analysis procedure, they are not considered in this example. In
general, these corrections should be provided by the experimental
collaborations as such an evaluation often requires access to
unpublished material. As described in Section~\ref{sec:acceptance},
the \eft code is prepared to include such correlations.

\subsection{Interpretation of measurements}

We assume no prior knowledge on the values of the four coupling
strengths, i.e. we choose the prior probability density for each
coupling to be uniform. The values of \gL and \gR are limited to a
range $[-1,1]$, the values of \VL and \VR are constrained to be within
$[-1.5, 1.5]$. These ranges are motivated by the fact that larger
anomalous couplings would have been observed by previous measurements.

\begin{figure}[t]
  \begin{center}
    \includegraphics[width=0.45\textwidth]{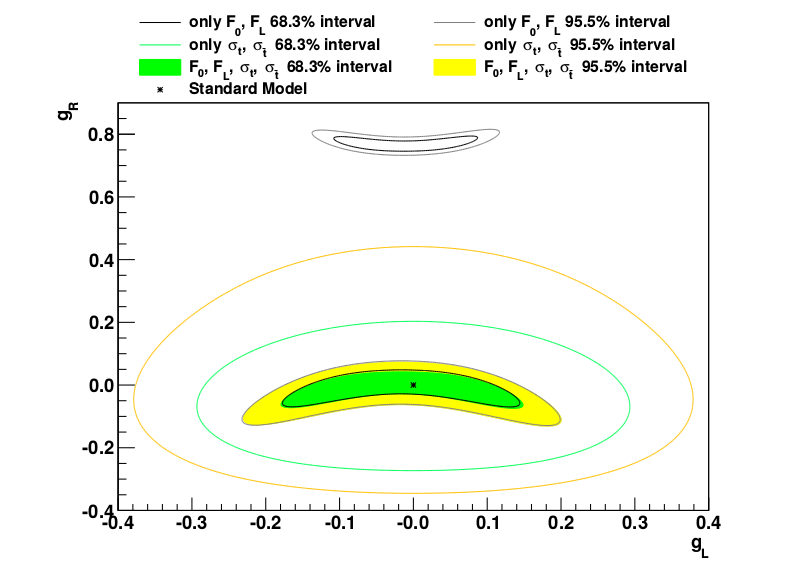} 
    \caption{Contours of the smallest areas containing 68.3\%, 95.5\%
      and 99.7\% posterior probability in the $(\gL, \gR)$-plane. The
      filled areas consider all four measurements, while the open ones
      take into account only two measurements. Also indicated is the
      SM prediction.
      \label{fig:2dfit}}
  \end{center}
\end{figure}

For the first interpretation, we assume SM values for the vector
couplings, i.e., we assume $\VL=1$ and $\VR=0$. Taking all four
measurements and their correlations into account,
Figure~\ref{fig:2dfit} shows the contours of the smallest areas
containing 68.3\% and 95.5\% probability in the two-dimensional plane
of \gR vs. \gL. In comparison, the dark and coloured lines indicate
these contours if only the measurements of the $W$-helicity or of the
$t$-channel cross sections are considered. While the measurement of
the $W$-helicity alone constrains two separate regions in
$(\gL,\gR)$-space, one centred around the SM prediction of $(0,0)$ and
another, smaller one around $(0, 0.8)$, the measurement of the
$t$-channel cross sections has less constraining power, but excludes
the second region. Using all four measurements thus reduces the
available parameter space for anomalous couplings by excluding the
second region and, if only marginally, by reducing the area of the
first region.

The relative importance of each of the four measurements is
illustrated in Tab.~\ref{tab:ranking}, which shows a ranking based on
the relative increase of the area within the 68.3\% contour if
individual measurements are ignored. This is compared to a ranking
based on the uncertainties of the one-dimensional marginal
distributions. As expected from Fig.~\ref{fig:2dfit}, the measurements
of the $W$-helicity have a larger impact on the constraints than the
$t$-channel measurements. It is worth noting, however, that the
ranking only addresses the size of the uncertainties, but not the
topology of the contours, i.e. the appearance of a second,
disconnected allowed region. The small negative values associated with
the measurement of $\sigma_{t}$ can be explained by the fact that the
measured value is furthest away from the SM prediction in comparison
to the other three measurements.

\begin{table}[ht!]
  \caption{Relative increase of the area contained in the 68.3\%
    posterior probability contour when removing one measurement at a
    time. Also indicated is the rank.
    \label{tab:ranking}}
  \begin{center}
    \begin{tabular}{llll}
      \hline
      Measurement & \multicolumn{3}{l}{Relative increase [\%] (rank)} \\ 
      & $(\gL, \gR)$ & \gL & \gR \\
      \hline 
      \fz              & \phantom{-}162.3 (2.) & \phantom{-}83.7 (1.) & \phantom{-}17.6 (2.) \\
      \fL              & \phantom{-}281.5 (1.) & \phantom{-}66.3 (2.) & \phantom{-}88.2 (1.) \\
      $\sigma_{\bar{t}}$ & \phantom{-00}2.2 (3.) & \phantom{-0}2.2 (3.) & \phantom{-0}0.0 (3.) \\
      $\sigma_{t}$      & -\phantom{00}3.2 (4.) & -\phantom{0}5.4 (4.) & -\phantom{0}2.9 (4.) \\
      \hline
    \end{tabular}
  \end{center}
\end{table}

\begin{figure}[t]
  \begin{center}
    \includegraphics[width=0.45\textwidth]{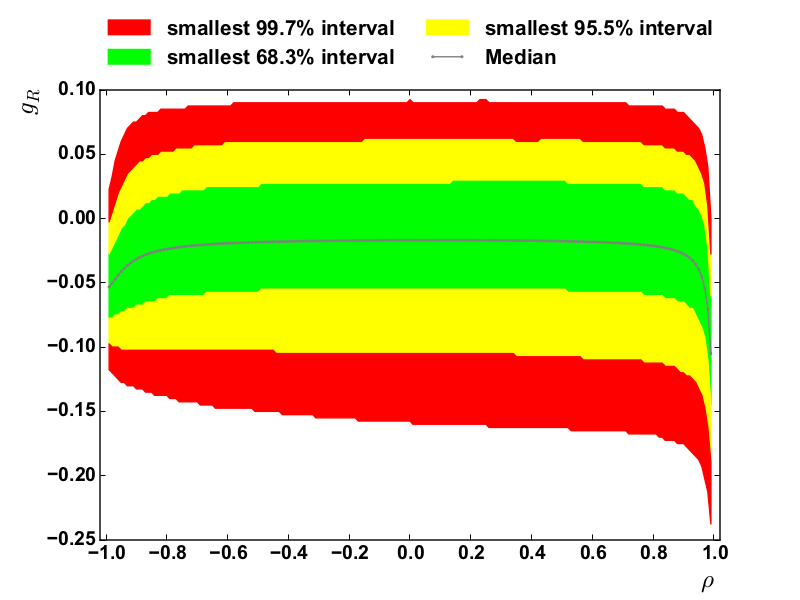} 
    \caption{Contours of the smallest areas containing 68.3\%, 95.5\%
      and 99.7\% posterior probability for \gR as a function of the
      linear correlation coefficient $\rho$ between the measurements
      of $\sigma_{t}$ and $\sigma_{\bar{t}}$.
      \label{fig:correlation}}
  \end{center}
\end{figure}

As an example, we test the impact of the correlation between
$\sigma_{t}$ and $\sigma_{\bar{t}}$ on the estimate of
\gR. Fig.~\ref{fig:correlation} shows the one-dimensional marginal
posterior probability for \gR as a function of the linear correlation
coefficient in the range $[-0.99,0.99]$. While the 68.3\%, 95.5\% and
99.7\% intervals do not vary significantly for values of $\rho$ in the
range $[-0.6, 0.6]$, they become smaller by up to a factor of two for
large, negative values and they become slightly larger for large,
positive values. Also, the median is shifted towards smaller values of
\gR in both extreme cases.

Instead of fixing the correlation coefficient between the two cross
section measurements, one can also assign an uncertainty to that
correlation. Assuming a Gaussian prior on the corresponding nuisance
parameter with a mean value of 0.5 and a standard deviation of 0.1,
the uncertainty on \gL and \gR does not change significantly. This is
expected from Fig.~\ref{fig:correlation} because the correlation does
not have a significant impact in this case.

\begin{figure}[t]
  \begin{center}
      \includegraphics[width=0.45\textwidth]{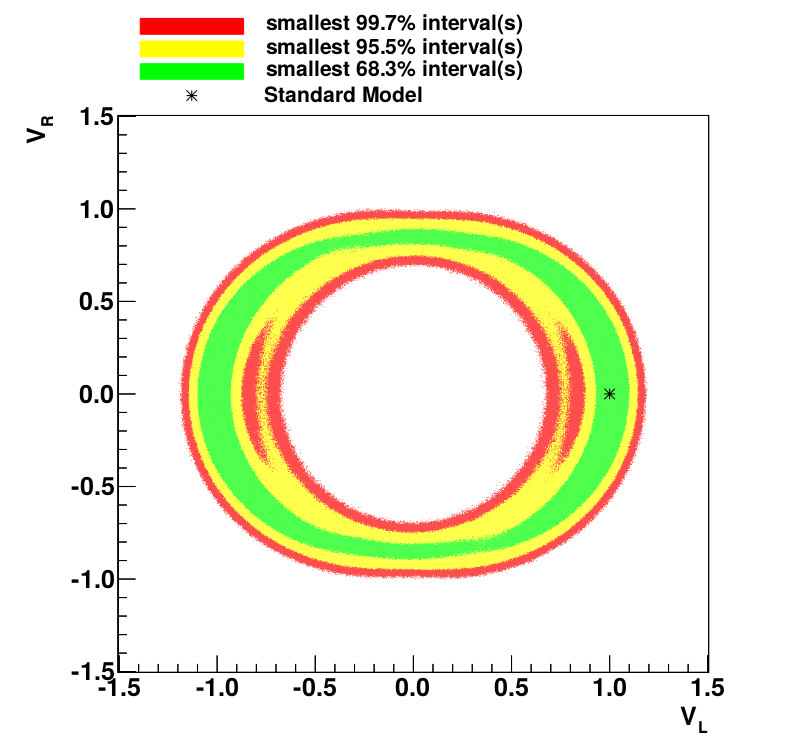} 
      \caption{Marginal posterior distributions in the $(\VL,
        \VR)$-plane when all four parameters are left free in the fit
        and all four measurements are considered.
        \label{fig:4dfita}}
  \end{center}
\end{figure}

\begin{figure}[t]
  \begin{center}
    \includegraphics[width=0.45\textwidth]{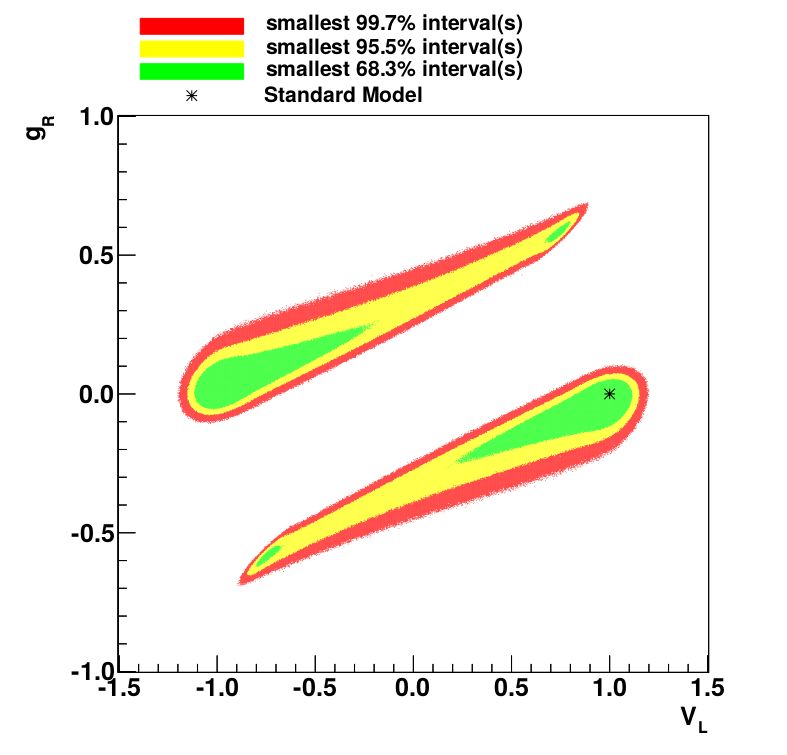} 
    \caption{Marginal posterior distributions in the $(\VL,
      \gR)$-plane when all four parameters are left free in the fit
      and all four measurements are considered.
      \label{fig:4dfitb}}
  \end{center}
\end{figure}

\begin{figure}[t]
  \begin{center}
      \includegraphics[width=0.45\textwidth]{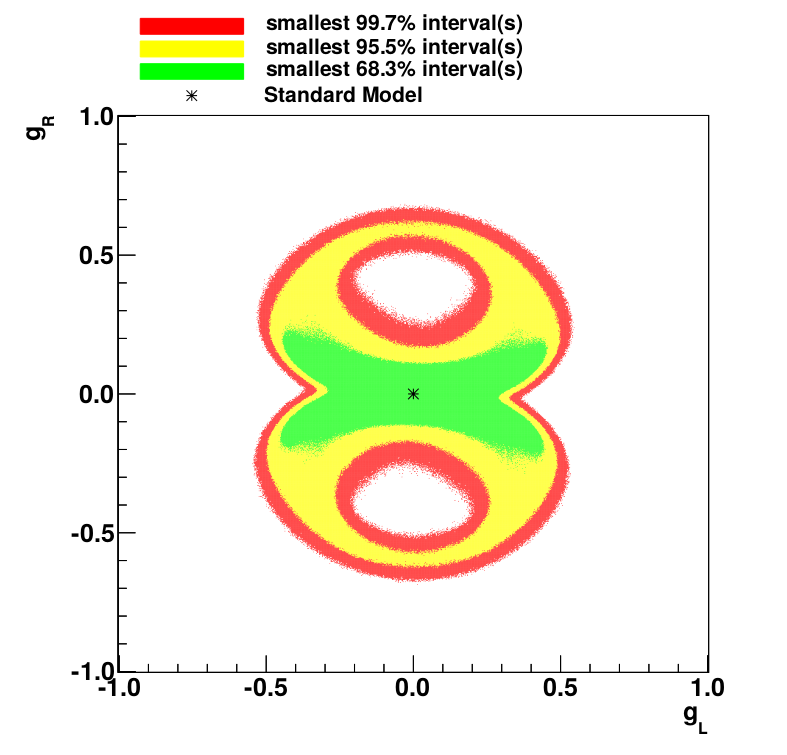} 
      \caption{Marginal posterior distributions in the
        $(\gL,\gR)$-plane when all four parameters are left free in
        the fit and all four measurements are considered.
        \label{fig:4dfitc}}
  \end{center}
\end{figure}

For the second interpretation, we assume all four couplings to be free
parameters of the fit. As an example, Figs.~\ref{fig:4dfita},
Fig.~\ref{fig:4dfitb} and Fig.~\ref{fig:4dfitc} show the marginal
posterior distributions in the $(\VL, \VR)$-plane, in the $(\VL,
\gR)$-plane, and in the $(\gL,\gR)$-plane, respectively. All three
distributions have highly non-Gaussian shapes and some show
disconnected regions. In each of the projections, the local mode is
consistent with the predictions of the SM and the absence of anomalous
couplings.


The global mode is found at $\VL=1.02$, $\VR=-0.06$, $\gL=0.03$, and
$\gR=-0.01$. Since the smallest four-dimensional hypervolume
containing 68.3\% posterior probability is strongly non-Gaussian in
shape and features several disconnected subsets, one-dimensional
measures, such as the standard deviation or smallest intervals, are
rendered useless.



\section{Conclusions}
\label{sec:conclusions}

We have presented a tool for interpreting measurements in the context
of effective field theories. This \eft allows implementing a
user-defined model, either directly or via interfaces to other
software tools, including predictions of observables based on the free
parameters of the model. Measurements of these observables are then
combined and used to constrain the free parameters. A variety of
features of the \eft helps to quantify and visualise the results. An
example in the field of top-quark physics was shown, for which
anomalous couplings of the $Wtb$-vertex were constrained based on
measurements of the $W$-boson helicity fractions and the single-top
$t$-channel cross sections.


\begin{acknowledgements}
The authors would like to thank Fabian Bach, Kathrin Becker, Dominic
Hirschb{\"u}hl and Mikolaj Misiak for their help and for the fruitful
discussions. In particular, the authors would like to thank Fabian
Bach for providing the code for the single-top cross
sections. N.C. acknowledges the support of FCT-Portugal through the
contract IF/00050/2013/CP1172/CT0002.
\end{acknowledgements}

\bibliographystyle{spphys}       
\bibliography{bibliography}   

\end{document}